\documentclass[10pt]{article}

\usepackage{amsmath}
\usepackage{amssymb}
\usepackage{graphics}
\usepackage{rotating}
\usepackage{cite}
\usepackage{color}
\usepackage{fancybox}
\usepackage{pstricks}

\def\0{\mbox{\tiny $0$}}
\def\1{\mbox{\tiny $1$}}
\def\2{\mbox{\tiny $2$}}
\def\3{\mbox{\tiny $3$}}
\def\4{\mbox{\tiny $4$}}
\def\5{\mbox{\tiny $5$}}
\def\6{\mbox{\tiny $6$}}
\def\7{\mbox{\tiny $7$}}
\def\8{\mbox{\tiny $8$}}
\def\9{\mbox{\tiny $9$}}
\title{\shadowbox{\large \bf \begin{tabular}{c}
RESONANCE, MULTIPLE DIFFUSION AND CRITICAL\\ TUNNELING FOR GAUSSIAN LASERS\end{tabular}}}

\author{
\small  Silv\^ania A. Carvalho\,\,\,\,and\,\,\,\,Stefano De
Leo\thanks{deleo@ime.unicamp.br} \\
\small Department of Applied Mathematics, State University of
Campinas, Brazil}

\date{\small
\fcolorbox{black}{yellow} {\color{red} $\bullet$ {\color{black}{
{\footnotesize  {\sc European Physical Journal D} {\bf 67}, 168-11 (2013)
}}} {\color{red}{$\bullet$}} } }

\begin{document}
%

\maketitle

\vspace*{-.7cm}

\begin{abstract}
\noindent We present a detailed study of the gaussian laser propagation through a dielectric system composed by two right angle prisms. We investigate the transition between the spatial coherence  limit, which exhibits wave-like properties and  for which the resonance phenomenon can be seen,
and the  decoherence limit, which exhibits particle-like properties and for which the
multiple diffusion occurs. We also analyze the tunneling at critical angles. In our numerical analysis, we shall use BK7 and Fused Silica prisms and  a gaussian $He$-$Ne$ laser with a wavelength of $632.8\,\mbox{nm}$ and  beam waists of $2\,\mbox{mm}$ and $200\,\mbox{$\mu$m}$.
\end{abstract}












\section*{\normalsize I. INTRODUCTION}

Total internal reflection in single right angle prism experiments has been the subject of a deep  study in the last decades\cite{ref11,ref12,ref13,ref14,ref15,ref16}. This phenomenon continues to be a topic of great scientific interest  because  an additional phase is present in the reflected beam when totally  reflected. This phase is responsible for the (Goos-H\"anchen) shift\cite{chen}  in the reflected beam. This shift is the optical counterpart of the well known delay time of quantum mechanics\cite{ref21,ref22,longhi}.

 The double right angle prism apparatus, see Fig.\,1, represents a fascinating arrangement to display both wave and particle-like propagation of light\cite{del1,del2} and to frustrate total internal reflection\cite{ref31,ref32,ref33,ref34,ref35,ref36,ref37}. In this paper, we shall use the wave packet formalism (gaussian beams) to describe the light propagation. The wave behavior is obviously recovered in the limit in which the beam waist is very large with respect to the air interspace between the two right angle prisms. The formalism used in this paper is the same presented in\cite{del2}, but,  with respect to the single dielectric block, the double prims apparatus present the additional phenomenon of frustrated total internal reflection, amplified  for incidence at {\em critical} angles. The use of wave packets  introduces a {\em new} parameter (the beam waist) in the numerical analysis. We investigate for
 BK7 and Fused Silica prisms and    gaussian $He$-$Ne$ lasers  how resonance, multiple diffusion and critical tunneling are related to this parameter. Of particular interest, it will be the study of  the limit case between wave and particle-like light propagation, i.e. the case of  {\em partial coherence}. The analytical and  numerical prediction shed new light on the transition between resonance and multiple diffusion and can be tested experimental.

\section*{\normalsize II. PARAXIAL APPROXIMATION AND GAUSSIAN LASER BEAM AMPLITUDE}

Gaussian beams are the simplest and often the most common type of beam
provided by a laser source. Their behavior is described by a few parameters
such as the spot size, the radius of curvature, and the Gouy phase\cite{born,saleh}.  In this
section, we introduce the wave number convolution of plane waves which, in the
limit of paraxial approximation, allows to obtain the standard analytic
expression for a gaussian beam moving in free space. This convolution function,
modulated by the reflection and transmission coefficients obtained by imposing continuity at each interface, will be then used to calculate the intensity of the outgoing  beams which propagate through the two right angle dielectric prisms separated by an air gap.

Our analysis is done for time-harmonic waves, $\omega(=2\,\pi c/\lambda)$. This
implies a stationary system. So, we can forego writing the time
dependence $\exp[-\,i\,\omega\,t\,]$. For our theoretical analysis, we start
with planes waves which correspond to momentum eigenvalues,
$\exp[\,i\,\boldsymbol{k}\cdot\boldsymbol{r}]$.  They determine a fixed
propagation direction $\boldsymbol{k}$. The electric filed can be then
expressed by $\exp[\,i\,\boldsymbol{k}\cdot\boldsymbol{r}
-\omega\,t\,]\,\,\boldsymbol{\hat{s}}$.
 The condition $\nabla \cdot \boldsymbol{E}=0$,
imposed by one of the Maxwell equations implies $\boldsymbol{\hat{s}}\cdot
\boldsymbol{k}=0$, and consequently the electric field vector lies in the
plane perpendicular to the direction of propagation $\boldsymbol{k}$.
Electromagnetic waves can oscillate with more than one orientation. By
convention, the polarization is described by specifying the orientation of the
electric field. For different  plane wave directions, the polarization vector
necessarily changes.

A general solution can then be obtained by convoluting plane waves with a
suitable integral
\begin{equation}
\boldsymbol{E}(\boldsymbol{r},t) =  E_{\0}\,\int\hspace*{-.1cm} \mbox{d}
\boldsymbol{k}\,\, G(k_x,k_y)\,\delta \left(k_{z} - \sqrt{k^{^{2}} -
k_{x}^{^2} - k_{y}^{^2}}\,\right)\,  \, \exp
\left[\,i\,(\,\boldsymbol{k}\cdot\boldsymbol{r}\,-\omega\,t\,)\,\right]
\,\,\boldsymbol{\hat{s}}\,\,,
\end{equation}
where $k=|\boldsymbol{k}|$. The Dirac delta function and the condition
$k=\omega/c$ guarantee  that each electric field component satisfies the wave
equation
\begin{equation}
\left(\,c^{\2}\,\nabla^{^{2}} -
\partial_t^{^{2}}\right)\,\boldsymbol{E}(\boldsymbol{r},t)
=0\,\,.
\end{equation}
For a sharp wave number distribution, $k_{x,y}\ll k$, the beam is sufficiently
collimated along the $z$ axis. Consequently,  we can use the following
approximation
\[k_z \approx k - \frac{k_x^{^2}+k_y^{^2}}{2\,k}\,\,.\]
Taking as plane of incidence the $y$-$z$ plane, we may assume to a good
approximation  that the electric vector is given by
$\boldsymbol{E}=\left\{\,E_s\,,\,E_p\,,\,0\,\right\}$, where $E_s$ is the
component of the electric field perpendicular to the plane of incidence
($s$-polarization) and $E_p$ is the component parallel to this plane
($p$-polarization). To simplify our presentation, we shall suppose that the
laser beam passes through a polarizer which selects the $s$-polarization,
\begin{eqnarray}
\boldsymbol{E}(\boldsymbol{r},t) &\approx & E_{\0}\,e^{\,i(kz-\omega
t)}\,\int\hspace*{-.1cm} \mbox{d}k_x\,\mbox{d}k_y\,\,\,\, G(k_x,k_y)\, \exp
\left[\,i\,\left(\,k_x\,x+k_y\,y - \frac{k_x^{^2}+k_y^{^2}}{2\,k}\,z\,\right)\,\right]
\,\,\boldsymbol{\hat{x}} \nonumber\\
 & = & E_{\0}\,e^{i(kz-\omega t)}\,A(x,y,z)\,\,\boldsymbol{\hat{x}}\,\,.
\end{eqnarray}
The field $A(x,y,z)$ satisfies the paraxial wave equation\cite{arf}
\begin{equation}
\left(\,\partial_{xx} + \partial_{yy} + 2\,i\,k\,\partial_z\,\right)\,A(x,y,z)
=0\,\,.
\end{equation}
Solving this equation yields an infinite set of functions of which the
gaussian beam is the lowest-order mode, \[ G(k_x,k_y) =
\frac{\mbox{w}_{\0}^{\2}}{4 \pi}\,\exp \left[-\frac{(k_{x}^{^2} +
k_{y}^{^2})\,\mbox{w}_{\0}^{\2}}{4}\right]\,\,.
\]
The use of the paraxial gaussian approximation is justified for beam waists,
$\mbox{w}_{\0}$, satisfying $k\,\mbox{w}_{\0}\geq 5$. From
\begin{equation}
A(x,y,z) =  \frac{\mbox{w}_{\0}^{\2}}{4 \pi}\,\, \int\hspace*{-.1cm}
\mbox{d}k_x\,\mbox{d}k_y\,\, \exp \left[-\frac{(k_{x}^{^2} +
k_{y}^{^2})\,\mbox{w}_{\0}^{\2}}{4}\right]\,  \, \exp
\left[\,i\,\left(k_x\,x+k_y\,y - \frac{k_{x}^{^2} +
k_{y}^{^2}}{2\,k}\,z\right)\right]\,\,,
\end{equation}
after performing two generalized gaussian integrations, we obtain the
well-known closed formula which describe the free propagation of a gaussian
laser beam
\begin{equation}
 E(x,y,z) =   E_{\0}\,e^{\,i(k\,z-\omega t)}\,\frac{\mbox{w}_{\0}}{\mbox{w}(z)}\,\,\exp \left[-\,\frac{x^{\2} +
y^{\2}}{\mbox{w}^{^{2}}(z)} + i\, \frac{(x^{\2} + y^{\2})\,k}{2\,R(z)}  - i
\phi(z) \right]\,\,,
\end{equation}
where
 \[\mbox{w}(z)=\mbox{w}_{\0}\,\sqrt{1+ \left(\frac{2\,z}{k\,\mbox{w}_{\0}^{^{2}}}\right)^{^{2}}}=
 \mbox{w}_{\0}\,\sqrt{1+ \left(\frac{z}{z_{_{R}}}\right)^{^{2}}}
  \]
  represents the radius at which the field amplitude drop to $1/e$ of its axial value,
  $z_{_{R}}$ is  the Rayleigh range,
 $R(z) = z\, (1+ z^{\2}_{_{R}}/z^{\2})$
is the radius of curvature of the wavefront of the gaussian beam, and
$\phi=\arctan[z/z_{_{R}}]$ is the longitudinal phase delay also know as  Gouy
phase. The corresponding intensity distribution is
 \begin{equation}
 I(x,y,z) =   I_{\0}\,\left[\frac{\mbox{w}_{\0}}{\mbox{w}(z)}\right]^{^{2}}\,\,\exp \left[-\,2\,\,\frac{x^{\2} +
y^{\2}}{\mbox{w}^{^{2}}(z)} \right]\,\,,
\end{equation}
and the total power of the beam is $P_{\0}=\pi\,\mbox{w}_{\0}^{\2}I_{\0}/2$.

\section*{\normalsize III. GEOMETRY OF THE DIELECTRIC SYSTEM}
In Fig.\,1, we show the $y$-$z$ overview of the double prism dielectric
layout. The double prism structure is created by air zones separating two right angle prisms made of the same dielectric material of refractive index $n$. The incident gaussian beam propagates along the $z$-axis,
\begin{equation}
\left[\, \nabla\,( \,\boldsymbol{k}\,\,\cdot \,\,\boldsymbol{r}) \,\right]_{_{(k_x=0,k_y=0)}} =   (\,0\,,\,0\,,\,k\,)\,\,,
\end{equation}
 and forms an angle $\theta$ with  $z_{_{up}}$, normal  to the dielectric block,
\begin{equation}
x_{_{up}}=x\,\,\,\,\,\,\,\mbox{and}\,\,\,\,\,\,\,\left( \begin{array}{c}
y_{_{up}} \\
z_{_{up}} \end{array} \right) = \left( \begin{array}{rr}
\cos\theta & \sin\theta \\
-\sin\theta & \cos\theta \end{array} \right)\left( \begin{array}{c}
y \\
z\end{array} \right)=  R\left(\theta\right) \left( \begin{array}{c}
y \\
z\end{array} \right)\,\,.
\end{equation}
Observing that $z_{_{up}}=z_{_{down}}$ and that  the two right angle prisms are separated by a  parallel air gap,  the down side outgoing beam has the same spatial phase of the incoming beam,
\[\boldsymbol{k_{_{down}}}\cdot \,\,\boldsymbol{r_{_{_{down}}}} =\,\,\,  \boldsymbol{k_{_{up}}}\cdot \,\,\boldsymbol{r_{_{_{up}}}} = \,\,\,\boldsymbol{k}\,\,\cdot \,\,\boldsymbol{r}\,\,.\]
For the right side outgoing beam the situation is a little bit more complicated. In this case, it is useful to introduce the following  rotations
\begin{equation}
x_{_{right}}=x_{*}=x_{_{up}}\,\,\,\,\,\,\,\mbox{and}\,\,\,\,\,\,\,
 \left( \begin{array}{c}
y_{_{right}} \\
z_{_{right}} \end{array} \right) =  R\left(-\frac{3\,\pi}{4}\right) \left(
\begin{array}{c}
y_{_{*}} \\
z_{_{*}} \end{array} \right) =  R\left(-\frac{\pi}{2}\right) \left(
\begin{array}{c}
y_{_{up}} \\
z_{_{up}} \end{array} \right)\,\,.
\end{equation}
Due to the fact that the first air/dielectric discontinuity is along $z_{_{up}}$ axis, we have
\[ q_{x_{up}}=k_{x_{up}}=k_x\,\,\,\,\,\,\,\mbox{and}\,\,\,\,\,\,\,
 q_{y_{up}}=k_{y_{up}}=k_y\cos\theta+ \sqrt{k^{^{2}} - k_x^{^{2}} - k_{y}^{^{2}}}\,\sin\theta\,\,.\]
Consequently, the spatial phase of the transmitted beam which propagates in the first dielectric block is given by
\begin{equation} \boldsymbol{q}_{_{up}}\cdot \,\,\boldsymbol{r_{_{_{up}}}} = \,k_x\, x+k_{y_{_{up}}}\, y_{_{up}} + \sqrt{n^{\2}k^{^{2}} - k_x^{^{2}} - k_{y_{_{up}}}^{^{2}}}\,\,z_{_{up}}
\,\,.\end{equation}
It is interesting to observe that for $k_{x,y}\to 0$, we find
\[ nk\,(\sin \psi,\cos \psi):=(q_{y_{up}},q_{z_{up}}) \to  k\,(\sin\theta,\sqrt{n^{\2}-\sin^{\2}\theta})\,\,\,\,\,\,\,\Rightarrow\,\,\,\,\,\,\,
\sin \theta= n\, \sin \psi\,\,.\]
Thus, the condition that the  wave number component perpendicular to the normal of the air/dielectric discontinuity   does not change, $q_{y_{up}}=k_{y_{up}}$, allows to recover, in the limit of plane wave,  the well known refraction Snell law.   The spatial phase of the beam which propagates from the transversal air/dielectric discontinuity towards the diagonal dielectric/air discontinuity in the first prism can be rewritten in terms of the $y_{_*}$-$z_{_*}$ axes as follows
\[  \boldsymbol{q_{_*}}\cdot \,\,\boldsymbol{r_{_*}}  =  \,\boldsymbol{q}_{_{up}}\cdot \,\,\boldsymbol{r_{_{_{up}}}}\,\,,\]
where
\[q_{x_{_*}}=q_{x_{_{up}}}=k_x\,\,\,\,\,\,\,\mbox{and}\,\,\,\,\,\,\,  \left( \begin{array}{c}
q_{y_{_*}} \\
q_{z_{_{*}}} \end{array} \right) =  R\left(\frac{\pi}{4}\right) \left(
\begin{array}{c}
q_{y_{_{up}}} \\
q_{z_{_{up}}} \end{array} \right)\,\,.   \]
The spatial phase of the reflected beam at the air gap is then obtained by replacing $q_{z_*}$ with $ -\, q_{z_*}$. Observing that
\[q_{x_{_{right}}}=q_{x_{_{*}}}=k_x\,\,\,\,\,\,\,\mbox{and}\,\,\,\,\,\,\,  \left( \begin{array}{c}
q_{y_{_{right}}} \\
q_{z_{_{right}}} \end{array} \right) =  R\left(-\,\frac{3\,\pi}{4}\right) \left(
\begin{array}{r}
q_{y_{_{*}}} \\
-\,q_{z_{_{*}}} \end{array} \right)=\left( \begin{array}{r}
-\,q_{y_{_{up}}} \\
q_{z_{_{up}}} \end{array} \right)=\left( \begin{array}{r}
-\,k_{y_{_{up}}} \\
q_{z_{_{up}}} \end{array} \right) \,\,,
\]
the reflected beam, propagating  from the diagonal dielectric/air interface to the right side dielectric/air discontinuity, will have  the following spatial phase
\begin{equation}
k_x\, x - k_{y_{_{up}}}\, y_{_{right}} + \sqrt{n^{\2}k^{^{2}} - k_x^{^{2}} - k_{y_{_{up}}}^{^{2}}}\,z_{_{right}}\,\,.
\end{equation}
Due to the fact that the right dielectric/discontinuity  is along the $z_{_{right}}$ axis, we have
\[ k_{x_{right}}=q_{x_{right}}=k_x\,\,\,\,\,\,\,\mbox{and}\,\,\,\,\,\,\,
 k_{y_{right}}=q_{y_{right}}=-\,k_{y_{_{up}}}\,\,.\]
Finally, the spatial phase of the right side outgoing beam is
\begin{eqnarray}
\boldsymbol{k_{_{right}}}\cdot \,\,\boldsymbol{r_{_{_{right}}}} & = & \,\,\,
k_x\, x+k_{y_{_{up}}}\, z_{_{up}} + \sqrt{k^{^{2}} - k_x^{^{2}} -
k_{y_{_{up}}}^{^{2}}}\,\,y_{_{up}} \nonumber \\
 & & k_x\,x \,+\,  \left[\,k_z\,\cos
(2\,\theta)-k_y\,\sin (2\,\theta)\,\right]\,y \,+\,\,
\left[\,k_z\,\sin (2\,\theta)+k_y\,\cos
(2\,\theta)\,\right]\,z\,\,,
 \end{eqnarray}
where $k_z=\sqrt{k^{^{2}}-k_x^{^{2}} - k_y^{^{2}}}$. The wave number vector of the right side outgoing beam is then given by
\begin{equation}
\left[\,\nabla\,\varphi_{_{right}}\,\right]_{_{(k_x=0,k_y=0)}} =   [\,0\,,\,k\,\cos(2\theta)\,,\,k\,\sin(2\,\theta)\,]\,\,.
\end{equation}
Whereas the down side outgoing beam and the incoming beam are always parallel, the right side outgoing beam and the incoming beam are parallel only for the incident angle $\theta=\pi/4$. Once obtained the spatial phases, to calculate the intensities of the outgoing beams, we need to find how the wave number distribution  is modified by reflection and transmission at each interface. For the right side outgoing amplitude, the convolution function is obtained by multiplying the incoming gaussian distribution by the transmission and reflection coefficients obtained by imposing  continuity at
the first transversal air/dielectric discontinuity, $T_{_{up}}$, at the air gap, $R_{_{*}}$, and at the (right) dielectric/air interface, $T_{_{right}}$,
\begin{equation}
A_{_{right}}(x,y,z) =  \frac{\mbox{w}_{\0}^{\2}}{4 \pi}\,\, \int\hspace*{-.1cm}
\mbox{d}k_x\,\mbox{d}k_y\,\,T_{_{up}}R_{_{*}}T_{_{right}} \exp \left[-\frac{(k_{x}^{^2} +
k_{y}^{^2})\,\mbox{w}_{\0}^{\2}}{4}\,+\,i\,\,\boldsymbol{k_{_{right}}}\cdot \,\,\boldsymbol{r_{_{_{right}}}}\,\right]\,\,.
\end{equation}
In a similar way, the down side outgoing amplitude is given by
\begin{equation}
\label{adown}
A_{_{down}}(x,y,z) =
  \frac{\mbox{w}_{\0}^{\2}}{4 \pi}\,\, \int\hspace*{-.1cm}
\mbox{d}k_x\,\mbox{d}k_y\,\,T_{_{up}}T_{_{*}}T_{_{down}} \exp \left[-\frac{(k_{x}^{^2} +
k_{y}^{^2})\,\mbox{w}_{\0}^{\2}}{4}\,+\,i\,\,\boldsymbol{k}\cdot \,\,\boldsymbol{r}\,\right]
\end{equation}
where $T_{_{*}}$ is the transmission amplitude for the beam propagating through the air gap
between the two right angle prisms and $T_{_{down}}$ is the coefficient for the transmission
through  the (down) dielectric/air interface.

\section*{\normalsize IV. REFLECTION AND TRANSMISSION COEFFICIENTS}

Maxwell´s equations describe optical phenomena and under certain conditions present a surprising aspect, they mimic the quantum mechanical Schr\"odinger equation\cite{cohen,grif}. The counterpart of the energy potential in quantum mechanics is represented by the stratified dielectric medium with which the laser interacts\cite{del1,longhi,del2,del3}. The reflection and transmission amplitudes both in optics  and quantum mechanics  can be built by imposing continuity at dielectric or potential discontinuities.

The incoming plane wave is
\[\exp[\,i\,\boldsymbol{k}\,\,\cdot \,\,\boldsymbol{r}\,]=\exp[\,i\, \boldsymbol{k_{_{up}}}\cdot \,\,\boldsymbol{r_{_{_{up}}}}]\,\,.
\]
For a dielectric block stratified along the $z_{_{up}}$ direction, the separation of variables implies that the wave number in the $x_{_{up}}(=x)$ and $y_{_{up}}$ directions remain unaltered, $q_{x_{_{up}}}=k_{x_{up}}=k_x$ and $q_{y_{_{up}}}=k_{y_{up}}$. Only the $z_{_{up}}$ wave number changes.  The continuity equations at the air/dielectric discontinuity
\[
\left[\,\exp[\,i\,k_{z_{_{up}}}z_{_{up}}] + R_{_{up}}\,\exp[\,-\,i\,k_{z_{_{up}}}z_{_{up}}]\,  \right]_{z_{_{up}}=d_{_{up}}} = \left[\, T_{_{up}}\,\exp[\,i\,q_{z_{_{up}}}z_{_{up}}] \, \right]_{z_{_{up}}=d_{_{up}}}
\]
and
\[
\left[\,\partial_{z_{_{up}}}\,\left(\,\exp[\,i\,k_{z_{_{up}}}z_{_{up}}] + R_{_{up}}\,\exp[\,-\,i\,k_{z_{_{up}}}z_{_{up}}]\right)  \right]_{z_{_{up}}=d_{_{up}}} = \left[\, \partial_{z_{_{up}}} T_{_{up}}\,\exp[\,i\,q_{z_{_{up}}}z_{_{up}}] \, \right]_{z_{_{up}}=d_{_{up}}}
\]
yield
\begin{equation}
T_{_{up}}  = \frac{2\,k_{z_{_{up}}}}{k_{z_{_{up}}} +\,
q_{z_{_{up}}}}\,\,\exp[\,i\,(k_{z_{_{up}}} - \,q_{z_{_{up}}})\, d_{_{up}}]\,\,.
\end{equation}
The transmission coefficients through the right and down dielectric interfaces can be immediately
obtained from the previous one by the following substitutions
\[ k_{z_{_{up}}}\to q_{z_{_{right,down}}}=q_{z_{_{up}}}\,\,\,,\,\,\,\,\,\,\, q_{z_{_{up}}}\to k_{z_{_{right,down}}}=k_{z_{_{up}}}
\,\,\,\,\,\,\,\mbox{and}\,\,\,\,\,\,\, d_{_{up}}\to d_{_{right,down}}\,\,.
\]
Consequently,
\begin{equation}
T_{_{right,down}}  = \frac{2\,q_{z_{_{up}}}}{q_{z_{_{up}}} +\,
k_{z_{_{up}}}}\,\,\exp[\,i\,(q_{z_{_{up}}} - k_{z_{_{up}}})\, d_{_{right,down}}]\,\,.
\end{equation}
The air gap between the right angle prisms simulates a quantum mechanical barrier of width $h_{_{*}}$. The general $z_{_{*}}$ solution in the air gap
\[ A_{_{*}}\,\exp[\,i\,k_{z_{_{*}}}z_{_{*}}] + B_{_{*}}\,\exp[\,-\,i\,k_{z_{_{*}}}z_{_{*}}] \]
can be oscillatory or evanescent depending if
\[k_{z_{_{*}}}^{^{2}} =  k^{^{2}} - k_x^{^{2}} - k_{y_{_{*}}}^{^{2}} \]
is positive or negative. Expanding  $k_{z_{_{*}}}^{^{2}}$ at first order in $k_{y}$,  we find
 \begin{equation}
 k_{z_{_{*}}}^{^{2}} =  k^{^{2}}\left(\,1 -\frac{\,\,n^{\2}}{2} - \sin\theta\,\sqrt{n^{^{2}}-\sin^{\2}\theta}\,\right) - k\,k_{y}\,\cos\theta\,\, \frac{n^{\2}-2\,\sin^{\2}\theta}{\sqrt{n^{^{2}}-\sin^{\2}\theta}} \,\,+\, \mbox{O}\left[k_x^{^{2}},k_y^{^{2}}\right]\,\,.
 \end{equation}
 Consequently, we find  evanescent waves for $k_y>\sigma(n,\theta)\,k$ and oscillatory waves for $k_y<\sigma(n,\theta)\,k$, where
 \begin{equation} \sigma(n,\theta) = \frac{2 - n^{\2} - 2\,\sin\theta\,\sqrt{n^{^{2}}-\sin^{\2}\theta}}{2\, \cos\theta\, (n^{\2}-2\,\sin^{\2}\theta)}\, \,\sqrt{n^{^{2}}-\sin^{\2}\theta}\,\,.
 \end{equation}
 At critical angles,
 \begin{equation}
\sin\theta_{_{c}}\,\sqrt{n^{\2} - \sin^{\2}\theta_{_{c}}}=1-\frac{\,\,n^{\2}}{2}\,\,,
\end{equation}
$\sigma(n,\theta) = 0$ and our wave packet will have the same amount of
evanescent ($k_y>0$) and oscillatory ($k_y<0$) waves.
 By imposing continuity of the electric field and its derivative at the air gap discontinuities, $z_{_{*}}=d_{_{*}}$ and $z_{_{*}}=d_{_{*}} + h_{_{*}}$, we find
\begin{equation}
\label{RTstar}
R_{_{*}} = -\,i\,\displaystyle{\frac{q^{\2}_{z_*} - k^{\2}_{z_*}}{2\,q_{z_*}\,k_{z_*}}}\, \frac{\sin(k_{z_*} h_{*})}{\mathcal{D_{_{*}}}}\, \,\exp[\,2\,i\,q_{z_{_{*}}}d_{_{*}}]\,\,\,\,\,\,\,\,\,\,\mbox{and}\,\,\,\,\,\,\,\,\,\,  T_{_{*}} =  \frac{\exp[-\,i\,q_{z_*} h_{*}]}{\mathcal{D_{_{*}}}}\,\,,
\end{equation}
where
\[
\mathcal{D_{_{*}}} = \cos(k_{z_*} h_{*}) - i\, \frac{q^{\2}_{z_*} +
k^{\2}_{z_*}}{2\,q_{z_*}\,k_{z_*}}\, \sin(k_{z_*} h_{*})\,\,.
\]

\section*{\normalsize V. RESONANCE AND MULTIPLE DIFFUSION}
In order to understand the transition between the coherence (wave-like behavior)  and decoherence (particle-like behavior) limit, let us observe that, for {\em oscillatory waves} propagating across the air gap,  the reflection and transmission coefficients given in Eq.\,(\ref{RTstar}) can be decomposed   in terms of multiple reflection and transmission terms\cite{del2} (see refs.\cite{ste1,ste2} for the quantum mechanical analog)
\begin{equation}
\label{incoherent}
R_{_{*}} =R_{_{*}}^{^{(1)}} + T_{_{*}}^{^{(1)}} R_{_{*}}^{^{(2)}} \sum_{_{m=0}}^{^{\infty}}\,\left(
\widetilde{R}_{_{*}}^{^{\,(1)}} R_{_{*}}^{^{(2)}} \right)^{^{m}}\, \widetilde{T}_{_{*}}^{^{\,(1)}} \,\,\,\,\,\,\,\mbox{and}\,\,\,\,\,\,\, T_{_{*}} =T_{_{*}}^{^{(1)}}  \sum_{_{m=0}}^{^{\infty}}\,\left(
\widetilde{R}_{_{*}}^{^{\,(1)}} R_{_{*}}^{^{(2)}} \right)^{^{m}}\, T_{_{*}}^{^{(2)}}\,\,,
\end{equation}
where
\[ R_{_{*}}^{^{(1)}} = \frac{q_{z_{_{*}}}-k_{z_{_{*}}} }{q_{z_{_{*}}} +\,
k_{z_{_{*}}}}\,\,\exp[\,2\,i\,q_{z_{_{*}}}\, d_{_{*}}]\,\, \,\,\,\,\,\mbox{and}\,\,\,\,\,\,\,
 T_{_{*}}^{^{(1)}} = \frac{2\,q_{z_{_{*}}}}{q_{z_{_{*}}} +\,
k_{z_{_{*}}}}\,\,\exp[\,i\,(q_{z_{_{*}}}-k_{z_{_{*}}})\, d_{_{*}}] \]
are the reflection and transmission coefficients for the wave which moving through the first dielectric block encounters  the  dielectric/air interface at $z_{_{*}}=d_{_{*}}$,
\[ R_{_{*}}^{^{(2)}} = \frac{k_{z_{_{*}}}-q_{z_{_{*}}} }{k_{z_{_{*}}} +\,
q_{z_{_{*}}}}\,\,\exp[\,2\,i\,k_{z_{_{*}}}\, (d_{_{*}}+h_{_{*}})]\,\, \,\,\,\,\,\mbox{and}\,\,\,\,\,\,\,
 T_{_{*}}^{^{(2)}} = \frac{2\,k_{z_{_{*}}}}{k_{z_{_{*}}} +\,
q_{z_{_{*}}}}\,\,\exp[\,i\,(k_{z_{_{*}}}-q_{z_{_{*}}})\, (d_{_{*}}+ h_{_{*}})] \]
are the reflection and transmission coefficients for the wave which moving across the air interspace  reaches the  second dielectric block  at $z_{_{*}}=d_{_{*}}+h_{_{*}} $, and
\[ \widetilde{R}_{_{*}}^{^{\,(1)}} = \frac{k_{z_{_{*}}}-q_{z_{_{*}}} }{k_{z_{_{*}}} +\,
q_{z_{_{*}}}}\,\,\exp[\,-\,2\,i\,k_{z_{_{*}}}\, d_{_{*}}]\,\, \,\,\,\,\,\mbox{and}\,\,\,\,\,\,\,
 \widetilde{T}_{_{*}}^{^{\,(1)}} = \frac{2\,k_{z_{_{*}}}}{k_{z_{_{*}}} +\,
q_{z_{_{*}}}}\,\,\exp[\,i\,(q_{z_{_{*}}}-k_{z_{_{*}}})\, d_{_{*}}]\,\,, \]
are the reflection and transmission coefficients for the wave which, reaching the second dielectric block,  propagating backwards to the first dielectric block  interface at $z_{_{*}}=d_{_{*}}$.

The phase difference between two successive contributions to reflection or transmission is given by the phase of the loop factor $\widetilde{R}_{_{*}}^{^{\,(1)}} R_{_{*}}^{^{(2)}}$. By using the stationary phase method, we can estimate the shift between two successive reflected or transmitted beams, i.e.
\[
2\,\left[\,\frac{\partial k_{z_{_{*}}}}{\partial k_y}
\right]_{_{(0,0)}}\hspace*{-.3cm}h_{_{*}} =\frac{\left(\,2\,\sin^{\2}\theta-n^{\2}\,\right) \cos \theta\,\sqrt{2} }{\sqrt{n^{^2}-\sin^{\2}\theta} \,\, \sqrt{2-n^2-2\,\sin \theta \, \sqrt{n^{^2}-\sin^{\2}\theta} }}\,\,h_{_{*}}\,\,.
\]
For $h_{_{*}}\ll \mbox{w}_{\0}$ (plane wave limit), the contributions in Eq.(\ref{incoherent}) overlap and  the beam manifests wave-like properties and  coherent  interference. In such a limit,  the relative power for the down/right-side outgoing beams is given by
\begin{eqnarray}
\label{cohP}
\frac{P_{_{down}}^{^{[wav]}}}{P_{_{0}}} & = & \left|\,T_{_{up}}T_{_{*}}T_{_{down}}\,\right|^{^{2}}_{_{(0,0)}} = \frac{16\,\cos^{\2}\theta\,(n^{\2}-\sin^{\2}\theta)}{\left(\, \cos\theta + \sqrt{n^{\2}-\sin^{\2}\theta}\,\right)^{^{4}}\,\left\{1+ \displaystyle{\frac{(n^{\2}-1)^{^{2}}\,\,\sin^{\2}\left[\,
f(n,\theta)\,k\,h_*\right]}{4\,\left[ n^{\2}-1+f^{^{2}}(n,\theta)\right] f^{^{2}}(n,\theta)}}\right\}}\,\,, \nonumber\\
\frac{P_{_{right}}^{^{[wav]}}}{P_{_{0}}}      & = & \left|\,T_{_{up}}R_{_{*}}T_{_{right}}\,\right|^{^{2}}_{_{(0,0)}} = \left|\,T_{_{up}}R_{_{*}}T_{_{down}}\,\right|^{^{2}}_{_{(0,0)}} = \frac{16\,\cos^{\2}\theta\,(n^{\2}-\sin^{\2}\theta)}{\left(\, \cos\theta + \sqrt{n^{\2}-\sin^{\2}\theta}\,\right)^{^{4}}}\,\, - \frac{P_{_{down}}^{^{[wav]}}}{P_{_{0}}} \,\,,
\end{eqnarray}
where
\[
f(n,\theta)=\sqrt{1-\displaystyle{\frac{\,\,n^{\2}}{2}}-\sin\theta \sqrt{n^{\2}-\sin^{\2}\theta}}\,\,.\]
For $\theta<\theta_{_{c}}$, $f(n,\theta)$ is a real positive number. Analysis of the behavior of the relative powers shows that resonances are found for
\begin{equation}
 \label{res}
 f(n,\theta_{_{m}})= m\,\frac{\pi}{k\,h_{_{*}}}= m\,\frac{\lambda}{\,2\,h_{_{*}}}\,\,,
 \end{equation}
where $m$ is a positive integer number. In Table 1, we give the first down-side outgoing beam resonance angles for BK7 and Fused Silica right angle prisms when the incident wavelength is $\lambda=632.8\,\mbox{nm}$. For small angles, a good approximation for $f(n,\theta_{_{m}})$  is given by
\[ f^{^{2}}(n,\theta_{_{m}})\approx 1-\frac{\,\,n^{\2}}{2}-n\,\theta_{_{m}}\,\,.\]
From Eq.(\ref{res}), we obtain
\begin{equation}
\theta_{_{m}} \approx  \frac{1}{n}\,\left[\,1-\left(m\,\frac{\lambda}{\,2\,h_{_{*}}}\right)^{^{2}}\,\right] -  \frac{n}{2}\,\,\,\,\,\,\,\,\,\,\Rightarrow\,\,\,\,\,\,\,\,\,\,\delta \theta_{_{m}}=\theta_{_{m+1}}-\theta_{_{m}}\approx -\,\frac{1+2\,m}{4\,n}\,\left( \frac{\lambda}{\,\,h_{_{*}}}\right)^{^{2}}\,\,.
\end{equation}
This clearly shows that,  for a fixed interval of incidence angles,  the numbers of the resonances increases for  increasing values of $h_{_{*}}$ and that, for a given air gap, the resonance width  increases for decreasing incidence angles. This behavior is also valid for the general case as it is clearly  shown in the power contour plots drawn in Fig.\,2.

Let us now analyze the case in which the contributions in Eq.(\ref{incoherent}) do {\em not} overlap, this happens for $h_{_{*}}$ greater than $\mbox{w}_{\0}$. In this case, incoherence reigns and the beams manifest a particle-like behavior. In such a limit, we have to consider the sum of the modulus squared of all contributions to reflection,
\begin{equation}
\left|\, R_{_{*}}\,\right|^{^{2}}_{_{[par]}} =\left|\,R_{_{*}}^{^{(1)}}\,\right|^{^{2}} +  \frac{\left|\,T_{_{*}}^{^{(1)}}R_{_{*}}^{^{(2)}} \widetilde{T}_{_{*}}^{^{\,(1)}} \,\right|^{^{2}} }{1-\left|\,\widetilde{R}_{_{*}}^{^{\,(1)}} R_{_{*}}^{^{(2)}}\,\right|^{^{2}} } = \frac{\left( q_{z_{_{*}}} - k_{z_{_{*}}} \right)^{^{2}} }{q_{z_{_{*}}}^{^{2}} +\,\, k_{z_{_{*}}}^{^{2}}}\,\,,
\end{equation}
and to transmission,
\begin{equation}
 \left|\,T_{_{*}}\,\right|^{^{2}}_{_{[par]}} = \frac{\left|\,T_{_{*}}^{^{(1)}}T_{_{*}}^{^{(2)}}\,\right|^{^{2}} }{1-\left|\,\widetilde{R}_{_{*}}^{^{\,(1)}} R_{_{*}}^{^{(2)}}\,\right|^{^{2}} } =
 \frac{2\, q_{z_{_{*}}}k_{z_{_{*}}}}{q_{z_{_{*}}}^{^{2}} +\,\, k_{z_{_{*}}}^{^{2}}}\,\,.
\end{equation}
The relative power for the down/right-side outgoing beams is now given by
\begin{eqnarray}
\label{incP}
\frac{P_{_{down}}^{^{[par]}}}{P_{\0}} & = &
\left[\,\left|T_{_{up}}\right|^{^{2}}\,\,\frac{2\, q_{z_{_{*}}}k_{z_{_{*}}}}{q_{z_{_{*}}}^{^{2}} +\,\, k_{z_{_{*}}}^{^{2}}}\,\,\left|T_{_{down}}\,\right|^{^{2}}\,\right]_{_{(0,0)}} \nonumber \\
 & = &
\frac{16\,\cos^{\2}\theta\,(n^{\2}-\sin^{\2}\theta)}{\left(\, \cos\theta + \sqrt{n^{\2}-\sin^{\2}\theta}\,\right)^{^{4}}} \,\,\, \frac{\left( \,  \sqrt{n^{\2}-\sin^{\2}\theta}-\sin\theta \,\right)\,\sqrt{2}\,\,f(n,\theta) }{1 - 2\,\sin\theta\, \sqrt{n^{\2}-\sin^{\2}\theta} }\,\,, \nonumber\\
\frac{P_{_{right}}^{^{[par]}}}{P_{\0}}           & = &
\left[\,\left|T_{_{up}}\right|^{^{2}}\,\,\frac{\left( q_{z_{_{*}}} - k_{z_{_{*}}} \right)^{^{2}} }{q_{z_{_{*}}}^{^{2}} +\,\, k_{z_{_{*}}}^{^{2}}}\,\,\left|T_{_{right}}\,\right|^{^{2}}\,\right]_{_{(0,0)}} =  \frac{16\,\cos^{\2}\theta\,(n^{\2}-\sin^{\2}\theta)}{\left(\, \cos\theta + \sqrt{n^{\2}-\sin^{\2}\theta}\,\right)^{^{4}}}\,\, - \frac{P_{_{down}}^{^{[par]}}}{P_{_{0}}}
\,\,.
\end{eqnarray}
In the particle limit, the power of the outgoing beams, as it is expected, do {\em not} depend on the air gap distance between the right angle prisms. This is due to the fact that for $h_{_{*}}$ greater than $\mbox{w}_{\0}$ the air gap (quantum barrier in quantum mechanics) has to be treated as two
separated air interfaces (the two step limit of quantum mechanics). In Table 2, we give the power
of the down-side outgoing beam for different incidence angles.

We conclude this section observing that, for oscillatory waves, the amplitudes which appear in  Eqs.(\ref{RTstar}) when summed lead to the amplitudes of Eq.(\ref{incoherent}). In the wave packet formalism,  the gaussian beams, depending on the ratio between the air gap distance and their beam waists, will automatically take into account whether the single contributions overlap or do no overlap. The theoretical result obtained in this section will be tested in section VI by numerical calculations from which we can also see the transition between the wave and particle-like behavior.

\section*{\normalsize VI. CRITICAL TUNNELING}

For $\theta>\theta_{_{c}}$,
\begin{equation}
\theta_{_{c}} = \left\{ \begin{array}{ccl}
+\,\arcsin \sqrt{(n^{^{2}}/\,2\,)-\sqrt{n^{^{2}}-1}} & \hspace*{.5cm} & \mbox{for}\,\,n<\sqrt{2}\,\,,\\
-\, \arcsin \sqrt{(n^{^{2}}/\,2\,)-\sqrt{n^{^{2}}-1}} &  & \mbox{for}\,\,n> \sqrt{2}\,\,,
\end{array}
\right.
\end{equation}
due to  the fact that  in Eqs.(\ref{cohP}) we have
\[ \frac{\sin\left[\,
f(n,\theta)\,k\,h_*\right]}{f(n,\theta)}  =\frac{\sin\left[\,i\,
|f(n,\theta)|\,k\,h_*\right]}{i\,|f(n,\theta)|} =\frac{\sinh\left[\,
|f(n,\theta)|\,k\,h_*\right]}{|f(n,\theta)|}\,\,,
\]
an exponential decay characterizes the transmission of light across the air gap.
For a fixed air gap (see Table 3), we experience, by changing the incidence angle $\theta$,  the phenomenon of frustrated total internal reflection and consequently tunneling of light. The transmission is maximized for incidence angles which approach to critical angles,  $f(n,\theta_{_{c}})\to 0$. For such angles, we find
\begin{eqnarray}
\label{cohP2}
\frac{P_{_{down}}^{^{[c]}}}{P_{\0}} & = & \frac{16\,\cos^{\2}\theta\,(n^{\2}-\sin^{\2}\theta)}{\left(\, \cos\theta + \sqrt{n^{\2}-\sin^{\2}\theta}\,\right)^{^{4}}\,\left[1+ \displaystyle{\frac{(n^{\2}-1)\,\,(k\,h_*)^{^{2}}}{4}}\right]}\,\,, \nonumber\\
\frac{P_{_{right}}^{^{[c]}}}{P_{\0}}            & = & \displaystyle{\frac{(n^{\2}-1)\,\,(k\,h_*)^{^{2}}}{4}} \,\,\frac{P_{_{down}}^{^{[c]}}}{P_{\0}} \,\,.
\end{eqnarray}
In Fig.\,3, we plot the relative power for the down/right-side outgoing $He$-$Ne$ beams. The evanescent waves propagate across the air gap and reaching  the second dielectric block allow
the phenomenon of frustrated total internal reflection. The contour plots show that this effect is amplified for decreasing values of the air gap distance between the right angle prisms and/or of the incidence angle. Consequently, experiments with incidence angles near to critical angles are more favorable to show the tunneling of light.

\section*{\normalsize VII. NUMERICAL ANALYSIS}

Up to now, we have presented analytical expressions for the relative power of the down/right-side outgoing beams.  In the case of diffusion, we have discussed two limits. The plane wave limit, see Eq.(\ref{cohP}), which we expect to be a good approximation for wave packets with beam waists $\mbox{w}_{\0} \gg h_{_*}$, and the particle limit,  see Eq.(\ref{incP}), which we expect to be valid for air gaps, $h_{_{*}}$ greater than the beam waists, $\mbox{w}_{\0}$. Obviously there is an intermediate situation which should reconciles these clear different limits. The transition between the wave and particle-like behavior is the subject matter of our numerical investigation. The numerical analysis will be also useful to test the theoretical prediction given in the previous sections.

We focus our attention on the power of the down-side outgoing beam. It is clear that a very similar analysis can be easily repeated for the right-side outgoing beam.  To calculate the power of the down-side outgoing beam, we have to numerically solve the following integral
\begin{equation}
P_{_{down}}  =  I_{\0}\,\,\int\hspace*{-.1cm}
\mbox{d}x\,\mbox{d}y\,\,\left|A_{_{down}}(x,y,z) \right|^{^{2}}\,\,.
\end{equation}
By using  the formula (\ref{adown}), given in section III, for $A_{_{down}}(x,y,z)$, the numerical calculation of the down-side outgoing beam power seems to require to solve six integrals,
\begin{eqnarray*}
P_{_{down}} & = &
  I_{\0}\,\,\left(\frac{\mbox{w}_{\0}^{\2}}{4 \pi}\right)^{^{2}}\,\, \int\hspace*{-.1cm}
\mbox{d}x\,\mbox{d}y\,\mbox{d}k_x\,\mbox{d}k_y\,\mbox{d}\tilde{k}_x\,\mbox{d}\tilde{k}_y\,\,
\times\nonumber \\
 & & \left[\,T_{_{up}}T_{_{*}}T_{_{down}}\,\right]_{_{(k_x,k_y)}} \exp \left[-\frac{(k_{x}^{^2} +
k_{y}^{^2})\,\mbox{w}_{\0}^{\2}}{4}\,+\,i\,\,\boldsymbol{k}\cdot \,\,\boldsymbol{r}\,\right]\,\,\times \nonumber \\
& & \overline{\left[\,T_{_{up}}T_{_{*}}T_{_{down}}\,\right]}_{_{(\tilde{k}_x,
\tilde{k}_y)}} \exp \left[-\frac{(\tilde{k}_{x}^{^2} +
\tilde{k}_{y}^{^2})\,\mbox{w}_{\0}^{\2}}{4}\,-\,i\,\,\boldsymbol{\tilde{k}}\cdot \,\,\boldsymbol{r}\,\right]\,\,.
\end{eqnarray*}
Nevertheless, the spatial integrations can be immediately done leading to two Dirac delta functions,
\[ (2\,\pi)^{^{2}}\,\, \delta(k_x-\tilde{k}_x)\,\,
\delta(k_y-\tilde{k}_y)\,\,.\]
Making use of these delta functions, we find
\begin{eqnarray}
P_{_{down}}
& = & I_{\0}\,\,\frac{\mbox{w}_{\0}^{\4}}{4}\,\, \int\hspace*{-.1cm}
\mbox{d}k_x\,\mbox{d}k_y\,\,
\left|\,T_{_{up}}T_{_{*}}T_{_{down}}\,\right|^{^{2}}_{_{(k_x,k_y)}} \exp \left[-\frac{(k_{x}^{^2} +
k_{y}^{^2})\,\mbox{w}_{\0}^{\2}}{2}\,\right]\,\,.
\end{eqnarray}
The relative power is then given by
\begin{equation}
\label{pdown1}
\frac{P_{_{down}}}{P_{\0}} =  \frac{\mbox{w}_{\0}^{\2}}{2\,\pi}\,\, \int\hspace*{-.1cm}
\mbox{d}k_x\,\mbox{d}k_y\,\,
\left|\,T_{_{up}}T_{_{*}}T_{_{down}}\,\right|^{^{2}}_{_{(k_x,k_y)}} \exp \left[-\frac{(k_{x}^{^2} +
k_{y}^{^2})\,\mbox{w}_{\0}^{\2}}{2}\,\right]\,\,.
\end{equation}
Observing that the motion is on the $y$-$z$ plane, the transmission coefficients are characterized by a linear dependence on $k_y$ and a quadratic dependence on $k_x$. Thus, for a narrow gaussian distribution (typical distribution for gaussian lasers) a good approximation for the relative power is given by
\begin{equation}
\label{pdown2}
\frac{P_{_{down}}}{P_{\0}} \approx  \frac{\mbox{w}_{\0}}{\sqrt{2\,\pi}}\,\, \int\hspace*{-.1cm}
\mbox{d}k_y\,\,
\left|\,T_{_{up}}T_{_{*}}T_{_{down}}\,\right|^{^{2}}_{_{(0,k_y)}} \exp \left[-\frac{
k_{y}^{^2}\,\mbox{w}_{\0}^{\2}}{2}\,\right]\,\,.
\end{equation}
In our numerical calculation, we use $He$-$Ne$ gaussian laser with $\lambda=632.8\,\mbox{nm}$ and beam waists of $200\,\mu\mbox{m}$ and $2\,\mbox{mm}$. This implies norrow distributions, $k\mbox{w}_{\0} \gg 1$. Consequently, the numerical analysis done by using Eq.(\ref{pdown2}) shows excellent agreement with the one done with  Eq.(\ref{pdown1}). In Fig.\,4 (BK7 right angle prisms) and Fig.\,5 (Fused Silica right angle prisms), we plot the relative power for the down-side outgoing beam for different values of the air gap interspace. For $h_{_*}\ll \mbox{w}_{\0}$, we recover the plane wave limit with the resonance phenomenon  predicted by Eq.(\ref{cohP}). For increasing values of the air gap interspace, the amplitude of the oscillations reduces reaching the particle limit where multiple diffusion occurs and where as it is predicted by  Eq.(\ref{incP}), we lost the dependence on the air gap distance between the dielectric blocks. In Figs.\,4 and 5, it is clear how the transition happens and how it is possible to reconcile the wave and particle-like behavior.  The numerical analysis also shows that the number of oscillations in a given interval is independent of the beam waist. The dependence on the beam waist is see in how the oscillations are damped.

In the case of  {\em partial coherence}, $h_{_{*}}\lesssim \mbox{w}_{\0}$, we can give a formula to estimate this damping. As mentioned before, for  $h_{_{*}}\ll \mbox{w}_{\0}$, we reproduce the plane wave limit and, consequently, the relative power for the down-side outgoing beams oscillates between a maximum and minimum value,
\begin{eqnarray}
\Delta[h_{_{*}}\ll\mbox{ w}_{\0}] & = &\frac{P_{_{down,\mbox{\tiny max}}}^{^{[wav]}}}{P_{_{0}}} - \frac{P_{_{down,\mbox{\tiny min}}}^{^{[wav]}}}{P_{_{0}}} =\left|\,T_{_{up}}T_{_{down}}\,\right|^{^{2}}_{_{(0,0)}}\left\{1-\left[ \frac{2\, q_{z_{_{*}}}k_{z_{_{*}}}}{q_{z_{_{*}}}^{^{2}} +\,\, k_{z_{_{*}}}^{^{2}}}\right]^{^{2}}_{_{(0,0)}}\right\}\nonumber \\
  & = & \left|\,T_{_{up}}T_{_{down}}\,\right|^{^{2}}_{_{(0,0)}}\left[ \frac{q_{z_{_{*}}}^{^{2}} -\,\, k_{z_{_{*}}}^{^{2}}}{q_{z_{_{*}}}^{^{2}} +\,\, k_{z_{_{*}}}^{^{2}}}\right]^{^{2}}_{_{(0,0)}}\,\,.
\end{eqnarray}
By using
\[ k_{z_{_{*}}}\approx \,k_{z_{_{*}}}^{^{(0,0)}}\, +\,\, \left[\,\frac{\partial k_{z_{_{*}}} }{\partial k_y}\,\right]_{_{(0,0)}}\hspace*{-.3cm}k_y\,\,,\]
we can develop the sine term as follows,
\begin{eqnarray*}
\sin(k_{z_{_{*}}}h_{_{*}}) &\approx&  \sin\left(k_{z_{_{*}}}^{^{(0,0)}}h_{_{*}}\right) \cos\left( \left[\,\frac{\partial k_{z_{_{*}}} }{\partial k_y}\,\right]_{_{(0,0)}}\hspace*{-.3cm}k_y h_{_{*}}\right) +
\cos\left(k_{z_{_{*}}}^{^{(0,0)}}h_{_{*}}\right) \sin\left( \left[\,\frac{\partial k_{z_{_{*}}} }{\partial k_y}\,\right]_{_{(0,0)}}\hspace*{-.3cm}k_y h_{_{*}}\right)
\nonumber \\
 &\approx  & \sin\left(k_{z_{_{*}}}^{^{(0,0)}}h_{_{*}}\right) \left\{ 1- \frac{1}{2}\, \left[\,\frac{\partial k_{z_{_{*}}} }{\partial k_y}\,\right]_{_{(0,0)}}^{^{2}}\hspace*{-.3cm}k^{^2}_y h_{_{*}}^{^{2}}\right\} +
\cos\left(k_{z_{_{*}}}^{^{(0,0)}}h_{_{*}}\right) \left[\,\frac{\partial k_{z_{_{*}}} }{\partial k_y}\,\right]_{_{(0,0)}}\hspace*{-.3cm}k_y h_{_{*}}\,\,.
  \end{eqnarray*}
From the previous equation, we get
  \begin{eqnarray}
\sin^{^2}(k_{z_{_{*}}}h_{_{*}})_{_{\mbox{\tiny max}}}
 &\approx  & \left[\,\frac{\partial k_{z_{_{*}}} }{\partial k_y}\,\right]_{_{(0,0)}}^{^{2}}\hspace*{-.3cm}k_y^{^{2}}h_{_{*}}^{^{2}}\,\,, \nonumber \\
 \sin^{^{2}}(k_{z_{_{*}}}h_{_{*}})_{_{\mbox{\tiny min}}}
 &\approx  &
 \left\{ 1- \frac{1}{2}\, \left[\,\frac{\partial k_{z_{_{*}}} }{\partial k_y}\,\right]_{_{(0,0)}}^{^{2}}\hspace*{-.3cm}k^{^2}_y  h_{_{*}}^{^{2}}\right\}^{^{2}} \approx
 \left\{ 1-  \left[\,\frac{\partial k_{z_{_{*}}} }{\partial k_y}\,\right]_{_{(0,0)}}^{^{2}}\hspace*{-.3cm}k^{^2}_y  h_{_{*}}^{^{2}}\right\}
 \,\,.
  \end{eqnarray}
Observing that $\langle k_y^{^{2}} \rangle = 1/2\,\mbox{w}_{\0}^{^{2}}$,
  \begin{eqnarray}
\delta[h_{_{*}}\lesssim\mbox{ w}_{\0}] & = &\left|\,T_{_{up}}T_{_{down}}\,\right|^{^{2}}_{_{(0,0)}}\left\{\frac{1}{1+\displaystyle{\left[ \frac{q_{z_{_{*}}}^{^{2}} -\,\, k_{z_{_{*}}}^{^{2}}}{2\, q_{z_{_{*}}}k_{z_{_{*}}}}\right]^{^{2}}_{_{(0,0)}}\left[\,\frac{\partial k_{z_{_{*}}} }{\partial k_y}\,\right]_{_{(0,0)}}^{^{2}}\frac{h_{_{*}}^{^{2}}}{2\,\mbox{w}_{\0}^{^{2}}}  }}\,\,-\right.
\nonumber \\
 & &  \left. \frac{1}{1+\displaystyle{\left[ \frac{q_{z_{_{*}}}^{^{2}} -\,\, k_{z_{_{*}}}^{^{2}}}{2\, q_{z_{_{*}}}k_{z_{_{*}}}}\right]^{^{2}}_{_{(0,0)}} \left(1 - \left[\,\frac{\partial k_{z_{_{*}}} }{\partial k_y}\,\right]_{_{(0,0)}}^{^{2}}\frac{h_{_{*}}^{^{2}}}{2\,\mbox{w}_{\0}^{^{2}}}\right)  }}\right\}\,\,.
\end{eqnarray}
Finally,
\begin{eqnarray}
\frac{\delta[h_{_{*}}   \lesssim  \mbox{ w}_{\0}]}{\Delta[h_{_{*}}\ll\mbox{ w}_{\0}]}
  & = & 1- \left\{\,2\, + \left[ \frac{\left(q_{z_{_{*}}}^{^{2}} -\,\, k_{z_{_{*}}}^{^{2}}\right)^{^{2}}}{2\,q_{z_{_{*}}} k_{z_{_{*}}} \left(q_{z_{_{*}}}^{^{2}} +\,\, k_{z_{_{*}}}^{^{2}}\right)}\right]^{^{2}}_{_{(0,0)}}\right\}\,\,\left[\,\frac{\partial k_{z_{_{*}}} }{\partial k_y}\,\right]_{_{(0,0)}}^{^{2}}\frac{h_{_{*}}^{^{2}}}{2\,\mbox{w}_{\0}^{^{2}}} \,\,.
\end{eqnarray}
This new formula shows as the plane wave resonance interval changes when $h_{_{*}}$ approaches the beam waist $\mbox{w}_{\0}$, i.e. in the limit of {\em partial coherence}.

\section*{\normalsize VIII. CONCLUSIONS}

In this paper, we have developed the mathematical formalism which allows to describe the propagation of gaussian lasers beam through a dielectric structure composed by two right angle prisms. Under certain conditions, $h_{_*}\ll \mbox{w}_{\0}$, the plane wave formulas, Eqs.(\ref{cohP}), represent a good approximation and the resonance phenomenon can be seen.  In this case the peaks of the incident and transmitted beams are clearly related. In the incoherence limit, when multiple diffusion appears, the relative power of the outgoing beam is not more dependent on the air gap interspace, see Eqs.\,(\ref{cohP}). In this case, only the first transmitted beam peak is related to the incoming beam. The secondary transmitted beams are related to the loop factor reflection given by the multiple diffusion of the  beam which propagates in the dielectric air gap. The numerical analysis confirms our analytical predictions for the wave and particle limit and shows the damped oscillation transition between these limits.

We hope that the material presented in our study  be useful to stimulate and prepare experimental setups to perform  qualitative and quantitative measurements to be used to confirm the results presented in our theoretical discussion of gaussian $He$-$Ne$ laser propagation through two dielectric (BK7 or Fused Silica) prisms. Due to the analogy between optics and quantum mechanics\cite{longhi}, the partial coherence limit could also play an important role in the study of transit time in periodic structure and consequently to shed new light on the discussion on superluminal transmission\cite{fin1,fin2,fin3}.

\section*{\small \rm ACKNOWLEDGEMENTS}

The authors thank the CNPq, grant PQ2/2013 (S.D.L), and
FAPESP, grant 2011/08409-0 (S.A.C) for financial support.
The author also thanks the referees for their suggestions and one of the referee
for drawing attention to the case of {\em partial} coherence which is discussed
at the end of section VII.


\newpage

\begin{center}
\begin{tabular}{|c|c|c|c||c|c|c|c|}
\hline
\multicolumn{8}{|c|}{{\bf Table 1}} \\ \hline  \hline
\multicolumn{8}{|c|}{$\lambda = 632.8\,\mbox{nm}$} \\ \hline \hline
\multicolumn{4}{|c||}{BK7 $[n=1.515]$} & \multicolumn{4}{|c|}{Fused Silica $[n=1.457]$} \\
\hline \hline
$h_{_{*}} [\mu \mbox{m}]$ & $\sin\theta_{\1}$ & $\sin\theta_{\2}$ & $\sin\theta_{\3}$ & $h_{_{*}} [\mu \mbox{m}]$ & $\sin\theta_{\1}$ & $\sin\theta_{\2}$ & $\sin\theta_{\3}$    \\ \hline
\hline
$1$ &  $-0.16457$ & $-0.37333$ & $-0.82546$ & $1$ &  $-0.11121$ & $-0.32521$ & $-0.78345$    \\ \hline
$2$ &  $-0.11436$ & $-0.16457$ & $-0.24960$ & $2$ &  $-0.05940$ & $-0.11121$ & $-0.19862$    \\ \hline
$5$ &  $-0.10038$ & $-0.10837$ & $-0.12170$ & $5$ &  $-0.04495$ & $-0.05321$ & $-0.06698$    \\ \hline
$10$ & $-0.09839$ & $-0.10038$ & $-0.10371$ & $10$ & $-0.04288$ & $-0.04495$ & $-0.04839$    \\ \hline
\end{tabular}
\end{center}
\noindent
Table 1: Wave-like behavior and coherent interference. First resonance angles for the down-side outgoing $He$-$Ne$ beam interacting with the dielectric structure composed by two BK7 and Fused Silica right angle prisms.

\vspace*{1cm}

\begin{center}
\begin{tabular}{|c|c||c|c|}
\hline
\multicolumn{4}{|c|}{{\bf Table 2}} \\ \hline  \hline
\multicolumn{1}{|c|}{$\theta[\frac{\pi}{12}]$} &  \multicolumn{1}{|c||}{$P_{_{down}}/\,P_{\0}$ - BK7} & \multicolumn{1}{|c|}{$\theta[\frac{\pi}{12}]$}  & \multicolumn{1}{|c|}{$P_{_{down}}/\,P_{\0}$ - Fused Silica} \\
\hline \hline
$-1$ & $0.62150$ & $-1$ & $0.70757$ \\ \hline
$-2$ & $0.74707$ & $-2$ & $0.79369$ \\ \hline
$-3$ & $0.73158$ & $-3$ & $0.77231$ \\ \hline
$-4$ & $0.61046$ & $-4$ & $0.65309$ \\ \hline
$-5$ & $0.32396$ & $-5$ & $0.35841$ \\ \hline
\end{tabular}
\end{center}
\noindent
Table 2: Particle-like behavior and multiple diffusion. Relative down-side outgoing beam  power  for the $He$-$Ne$ beam interacting with the dielectric structure composed by two BK7 and Fused Silica right angle prisms.

\vspace*{1cm}

\begin{center}
\begin{tabular}{|c|l|l|l|l||c|l|l|l|l|}
\hline
\multicolumn{10}{|c|}{{\bf Table 3}} \\ \hline  \hline
\multicolumn{10}{|c|}{$\lambda = 632.8\,\mbox{nm}$} \\ \hline \hline
\multicolumn{1}{|c|}{$\bullet$} &  \multicolumn{4}{|c||}{$P_{_{down}}/\,P_{\0}$ - BK7} & \multicolumn{1}{|c|}{$\bullet$} & \multicolumn{4}{|c|}{$P_{_{down}}/\,P_{\0}$ - Fused Silica} \\
\hline \hline
$h_{_{*}} [\mu \mbox{m}]$ & $\theta=\theta_{_c}$ & $\theta=-\,\frac{\pi}{36}$ & $\theta=-\,\frac{\pi}{60}$  & $\theta=0$ & $h_{_{*}} [\mu \mbox{m}]$ & $\theta=\theta_{_c}$ & $\theta=-\,\frac{\pi}{90}$ & $\theta=0$  & $\theta=\frac{\pi}{60}$
   \\ \hline
\hline
$1$ & $2.78\,[2]$ &  $1.69\,[2]$ & $4.09\,[3]$ & $7.19\,[4]$ &
$1$ & $3.25\,[2]$ &  $2.32\,[2]$ & $5.66\,[3]$ & $1.01\,[3]$
 \\ \hline
$2$ & $7.12\,[3]$ &  $1.20\,[3]$ & $2.25\,[5]$ & $3.48\,[7]$ &
$2$ & $8.34\,[3]$ &  $2.41\,[3]$ & $4.09\,[5]$ & $6.39\,[7]$
 \\ \hline
 $5$ & $1.15\,[3]$ &  $6.46\,[7]$ & $3.80\,[12]$ & $3.95\,[17]$ &
$5$ & $1.34\,[3]$ &  $5.03\,[6]$ & $1.58\,[11]$ & $1.61\,[16]$
 \\ \hline
 $10$ & $2.87\,[4]$ &  $2.34\,[12]$ & $1.97\,[23]$ & $1.05\,[33]$ &
$10$ & $3.37\,[4]$ &  $1.81\,[10]$ & $3.22\,[22]$ & $1.61\,[32]$
 \\ \hline
\end{tabular}
\end{center}
\noindent
Table 3: Tunneling of light. Relative down-side outgoing beam  power  for the $He$-$Ne$ beam interacting with the dielectric structure composed by two BK7 and Fused Silica right angle prisms.
In the table  $[num]$ stands for $10^{^{-num}}$. By decreasing the air gap distance between the right angle prisms and/or the incidence angle, we experience the phenomenon of frustrated total internal reflection.


\newpage

\begin{figure}
\vspace*{-5cm} \hspace*{-3.5cm}
\includegraphics[width=23cm, height=30cm, angle=0]{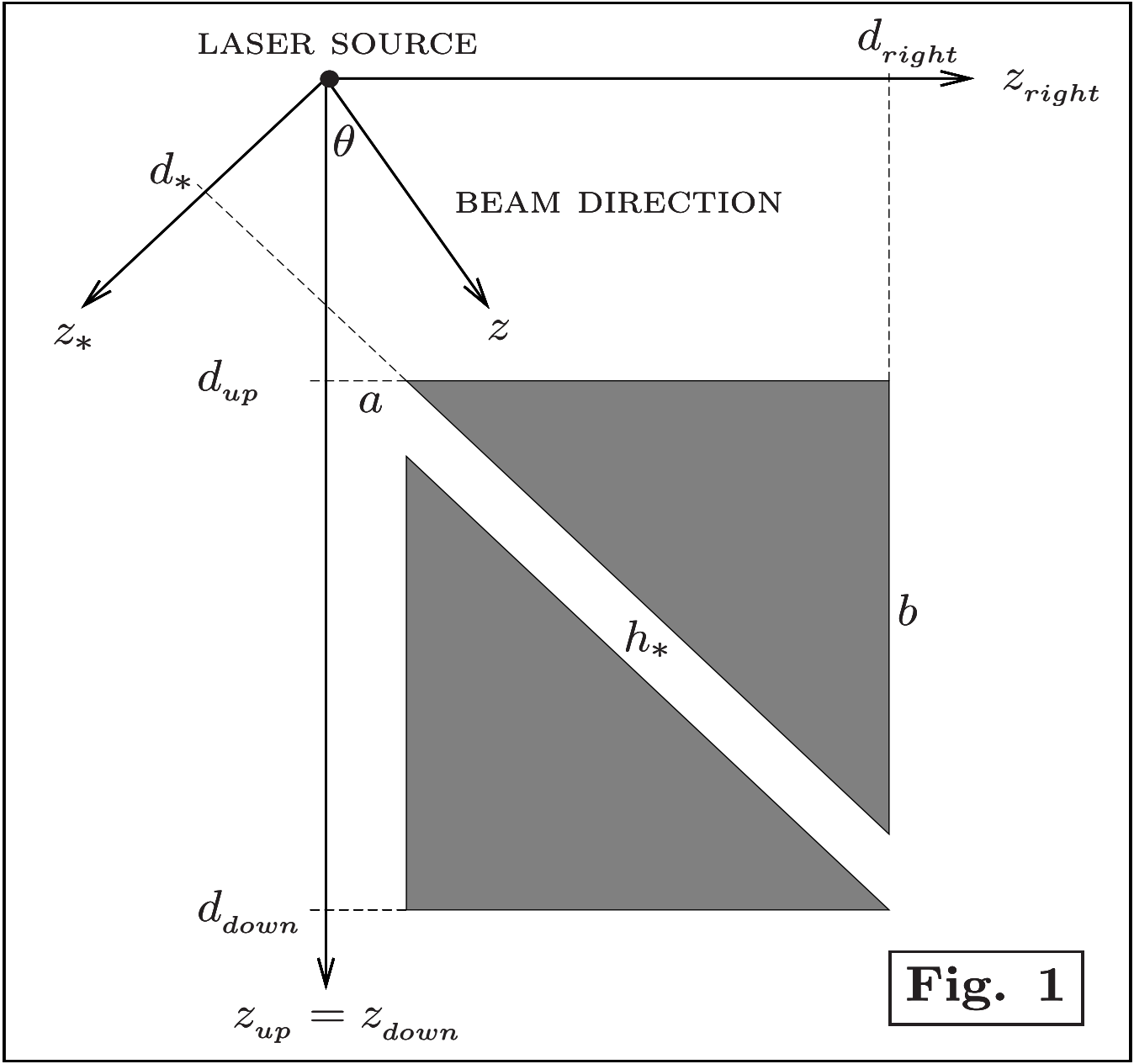}
\vspace*{-7.5cm}
 \caption{Geometric layout of the dielectric structure analyzed in this paper. The two right angle prisms are separated by a narrow air gap. Depending on the incidence angle, $\theta$, and on the air gap dimension, $h_{_{*}}$,   the power of the down/right-side outgoing beams will exhibit resonance, multiple diffusion and tunneling phenomena.}
\end{figure}

\newpage

\begin{figure}
\vspace*{-1.cm} \hspace*{-2.5cm}
\includegraphics[width=20cm, height=26cm, angle=0]{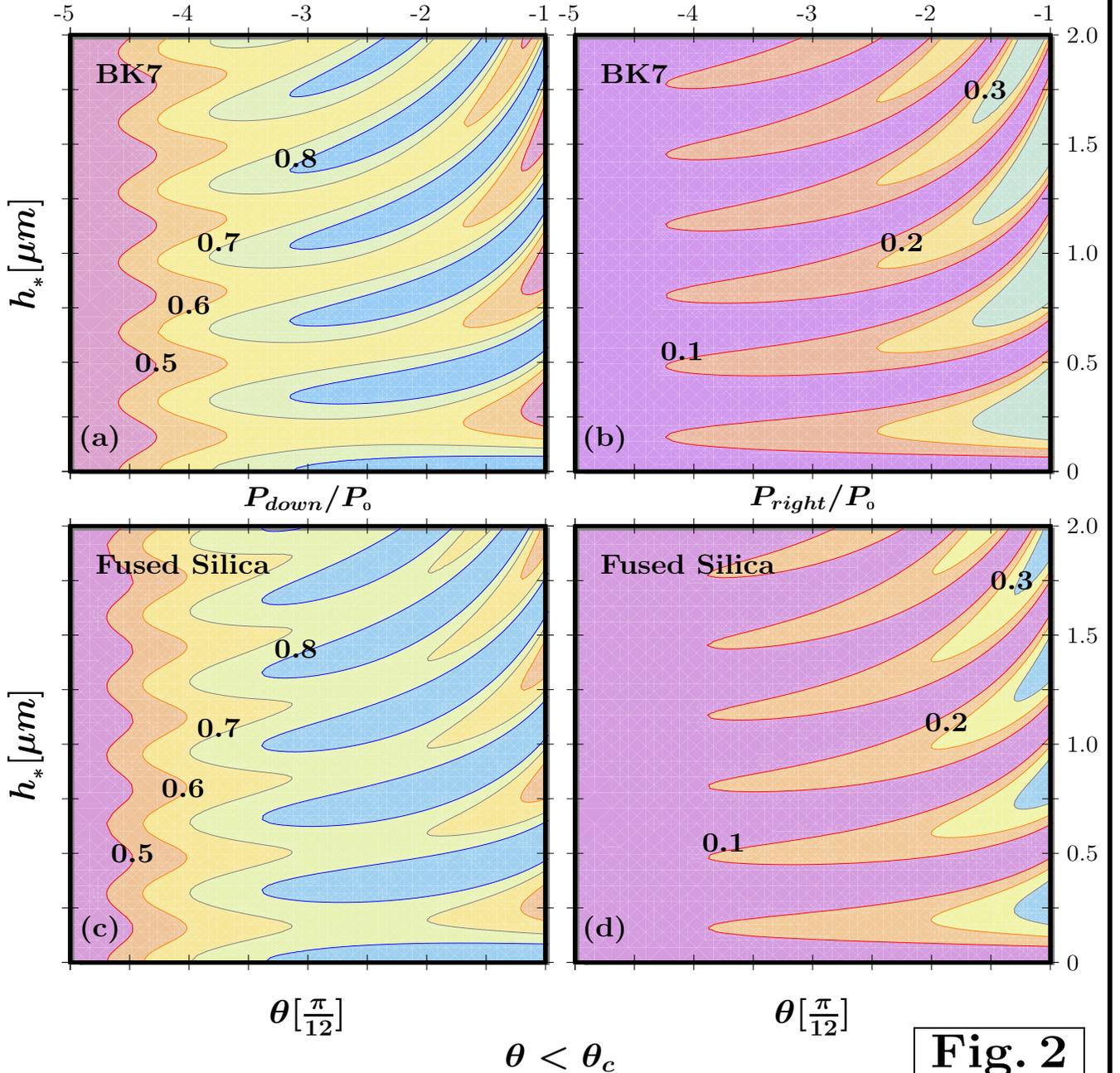}
\vspace*{-5cm}
 \caption{Relative power contour plots for the plane wave limit ($h_{_{*}}\ll \mbox{w}_{\0}$) for which wave-like properties can be seen. For $\theta<\theta_c$, the waves which propagate across the air gap are oscillatory and consequently the phenomenon of resonance appears. The contour plots show that for a fixed interval of incidence angles  the numbers of the resonances increase by increasing the air gap space between the two right angle prisms and that for given air gap the resonance width increases for decreasing incidence angle.  }
\end{figure}

\newpage

\begin{figure}
\vspace*{-1.cm} \hspace*{-2.5cm}
\includegraphics[width=20cm, height=26cm, angle=0]{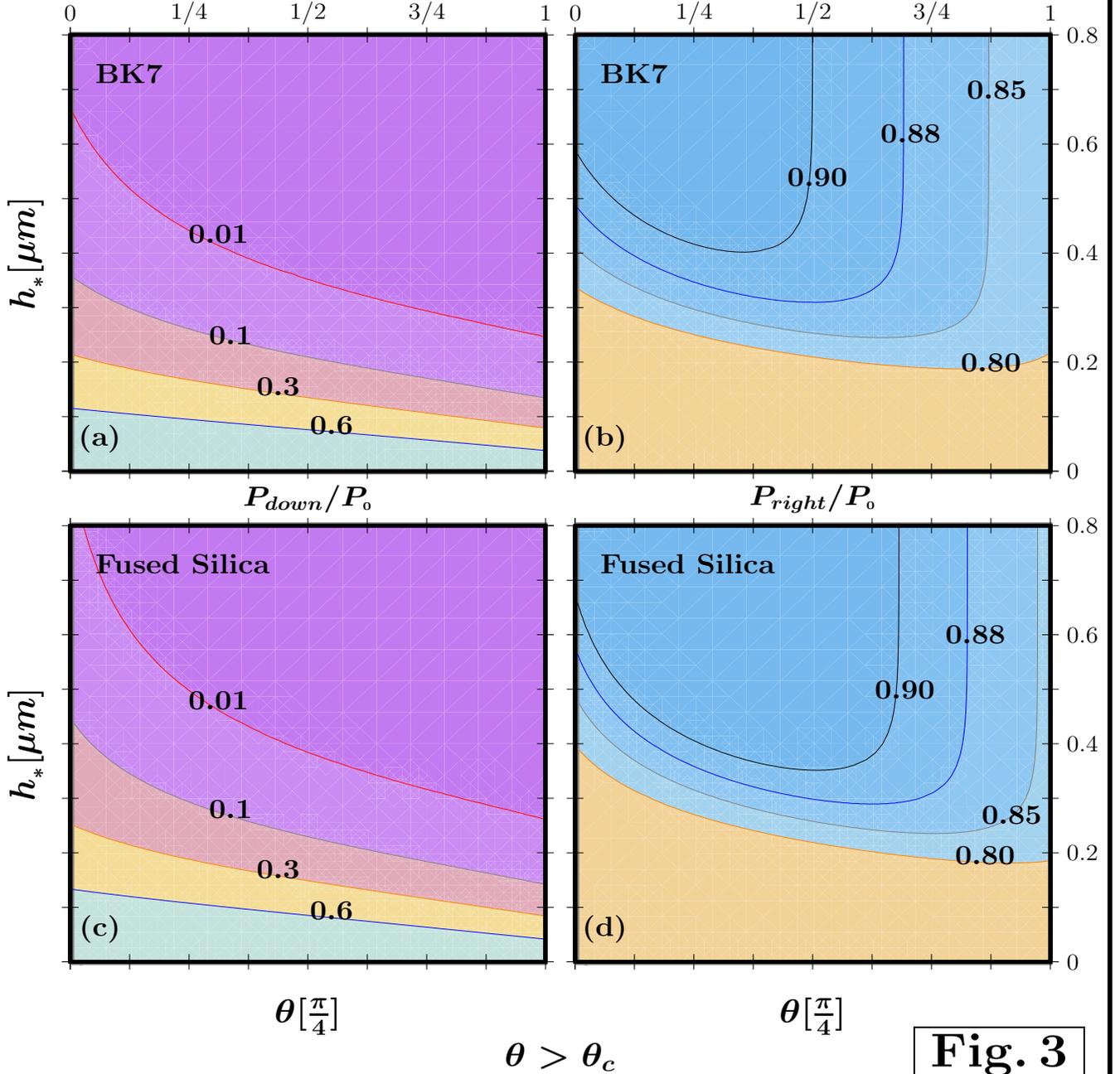}
\vspace*{-5cm}
 \caption{Relative power contour plots for tunneling. For $\theta>\theta_c$, the waves which propagate across the air gap are evanescent and consequently the phenomenon of frustrated total internal reflection appears for the down-side outgoing beam. The contour plots show that this effect is amplified for decreasing values of the air gap distance between the right angle prisms and/or of the incidence angle. From the contour plots is also clear that experiments with incidence angles near to critical angles are more favorable to show the phenomenon of frustrated total internal reflection.}
\end{figure}

\begin{figure}
\vspace*{-1.cm} \hspace*{-3.5cm}
\includegraphics[width=21cm, height=25cm, angle=0]{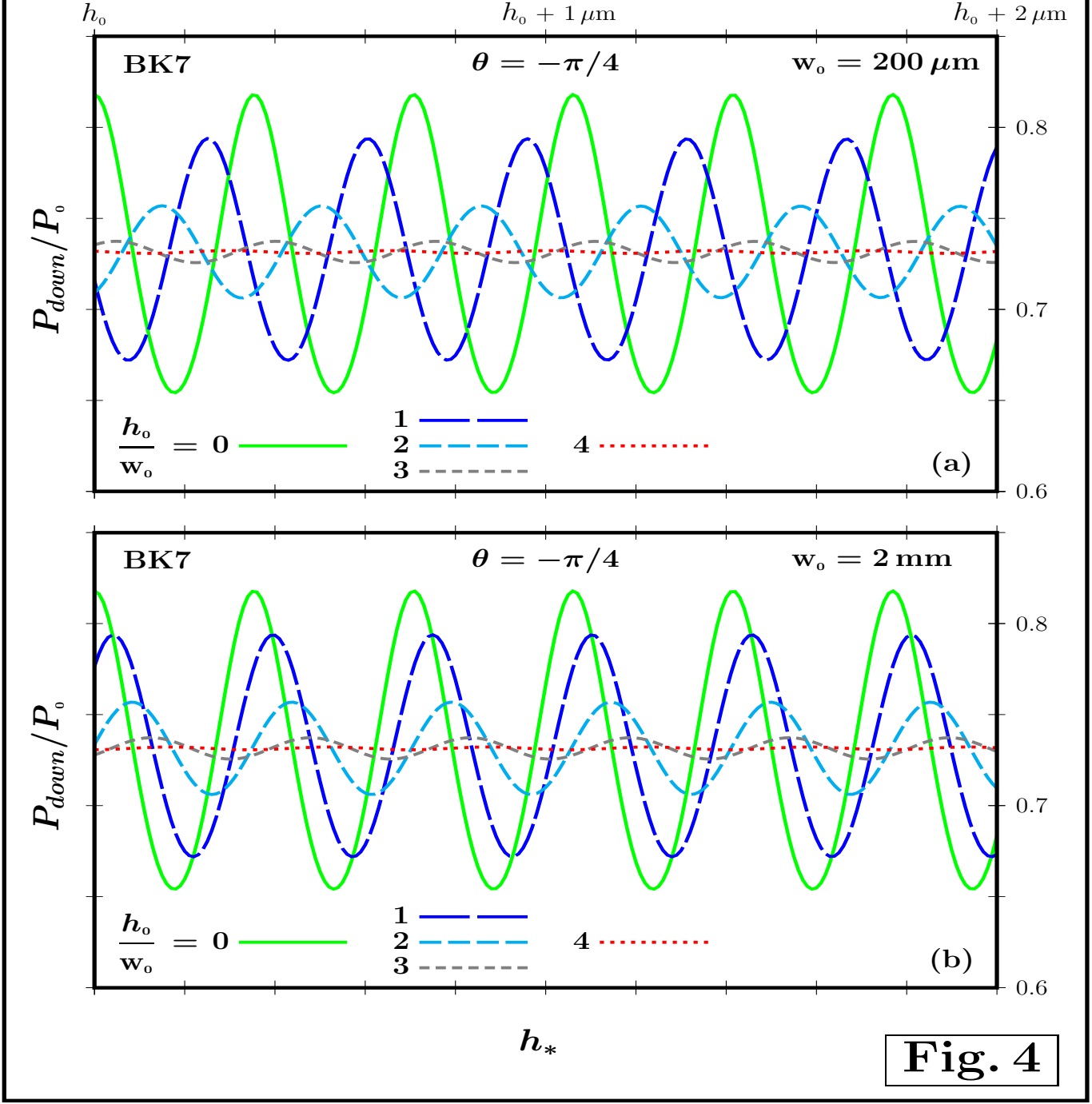}
\vspace*{-3.5cm}
 \caption{The relative power for the down-side outgoing beam propagating through BK7 right angle prisms is plotted for different values of the air gap which separates the dielectric blocks. For small values of $h_{_{*}}$ ($h_{\0}=0$), we find the typical resonance curves  of the plane wave analysis. For increasing values of $h_{_*}$ ($h_{\0}/\mbox{w}_{\0}=1,\,2,\,3$), the amplitude of the oscillations decreases approaching, for $h_{\0}=4\,\mbox{w}_{\0}$, the particle limit of Table 2 ($\approx 0.732$). The numerical analysis confirms our theoretical predictions.}
\end{figure}

\newpage

\begin{figure}
\vspace*{-1.cm} \hspace*{-3.5cm}
\includegraphics[width=21cm, height=25cm, angle=0]{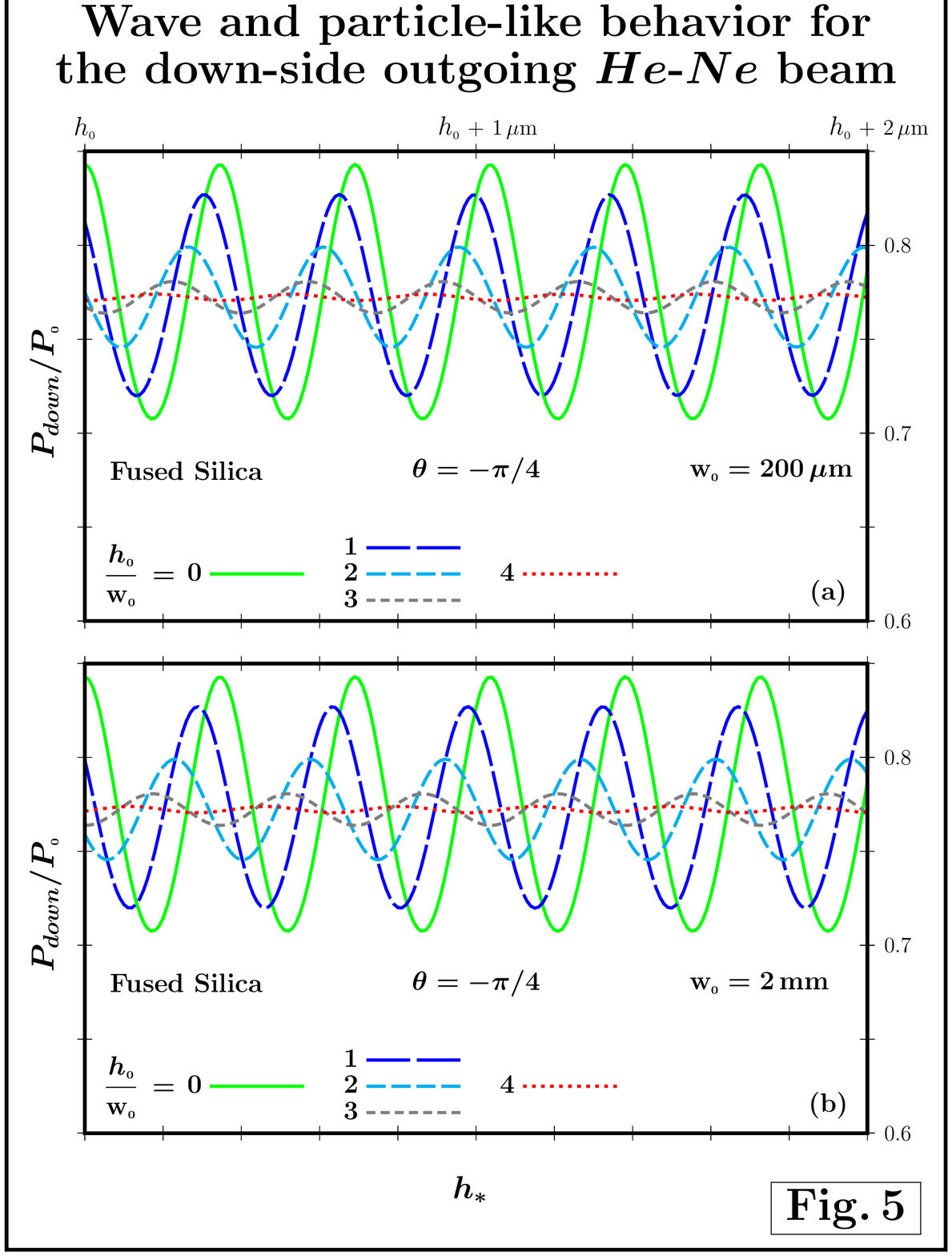}
\vspace*{-3.5cm}
 \caption{The relative power for the down-side outgoing beam propagating through Fused Silica right angle prisms is plotted for different values of the air gap which separates the dielectric blocks. For small values of $h_{_{*}}$ ($h_{\0}=0$), we find the typical resonance curves  of the plane wave analysis. For increasing values of $h_{_*}$ ($h_{\0}/\mbox{w}_{\0}=1,\,2,\,3$), the amplitude of the oscillations decreases approaching, for $h_{\0}=4\,\mbox{w}_{\0}$, the particle limit of Table 2 ($\approx 0.772$). The numerical analysis confirms our theoretical predictions.}
\end{figure}

\end{document}